\documentstyle[11pt,rafringe,twoside,epsfig,makeidx,graphicx]{article}

\makeindex

\def\edcomment#1{\iffalse\marginpar{\raggedright\sl#1\/}\else\relax\fi}
\reversemarginpar

\begin{document}
\title{Motion Statistics in the CJ Survey -- The Status in October 2002}
\author{R. C. Vermeulen}
\affil{Astron, P.O. Box 2, 7990 AA, Dwingeloo, The Netherlands}
\author{S. Britzen}
\affil{Astron, P.O. Box 2, 7990 AA, Dwingeloo, The Netherlands}
\affil{Landessternwarte, K\"onigstuhl, 69117 Heidelberg, Germany}
\author{G. B. Taylor}
\affil{National Radio Astronomy Observatory, PO Box O, Socorro, NM 87801, USA}
\author{T. J. Pearson, A. C. S. Readhead}
\affil{California Institute of Technology, Pasadena, CA 91125, USA}
\author{P. N. Wilkinson, I. W. A. Browne}
\affil{Jodrell Bank Observatory, University of Manchester, Macclesfield, Cheshire SK11 9DL, UK}

\begin{abstract}
In the Caltech-Jodrell Survey of bright flat-spectrum radio sources,
3--5 epochs have now been observed for nearly all 293~sources; a
uniquely large sample. The derivation of component motions has not yet
been completed; it is complicated by, for example, variability and jet
curvature. Nevertheless, some basic results are clear. The average
apparent velocity in CJF quasars is near $3c$, while for radio galaxies
and BL\,Lacs it is around $1c$. The distribution of velocities is broad,
indicating a broad distribution of jet Lorentz factors, with many low
values, and/or a decoupling of the Lorentz factors between the cores
and the moving jet components, due to bending, speed changes, or due to
pattern motions (shocks). A hint has emerged at this workshop that the
average apparent velocities might be lower at lower radio frequencies;
this will need careful verification.
\end{abstract}

\section{Introduction}

Superluminal motion is one of the many astronomical research topics in
which Ken Kellermann has been deeply involved ever since it arose. Ken
was of course already playing a central role in the exploration of
variability in extragalactic radio sources (e.g., Pauliny-Toth \&
Kellermann 1966; Kellermann 1966). Then, with the development of VLBI
(one of the earlier papers on the subject being Kellermann et al.\
1968), came the gradual establishment of the fact that, as predicted by
Rees (1967), evolving, relativistically moving components not only
cause rapid flux density variability, but indeed also display
superluminal velocities, due to geometrical effects when the motion is
directed close to our line of sight. It is historically interesting to
see the development of this interpretation in papers (co)authored by
Ken through the late 1960s, the 1970s, and even the early 1980s, for
example Cohen et al.\ (1971), Shaffer et al. (1972), Kellermann et
al. (1973), Kellermann et al.\ (1974), and Cohen et al. (1977);
perhaps this can be said to have culminated in Ken's 1985 review
(Kellermann 1985).

In the early days, there was a focus on some very interesting
individual cases of rapidly varying and very bright sources, including
quasars such as \Index{3C\,273} and \Index{3C\,279}, 
as well as galaxies such as \Index{3C\,84} and
\Index{3C\,120}. But, clearly, the statistics of a significantly sized sample are
needed in order to obtain a more systematic overview of jet properties,
such as the range of Lorentz gamma factors, and, for example, to probe
the occurrence of pattern motions distinct from the bulk flow (e.g.,
Vermeulen \& Cohen 1994). A more statistical approach is also needed to
study how the properties of relativistic jets are related to the other
physical parameters of the radio sources and their hosts, and also to
test unification models, which explain apparent differences between
object classes as being due to viewing them at different angles to the
line of sight. Ken himself long ago recognised the need for large
surveys, and, when the VLBA, of which he has of course been a long-term
advocate, started operations in the 1990s, he initiated the 2\,cm VLBA
survey (Kellermann et al.\ 1998), as described elsewhere in these
proceedings.

Meanwhile, the 6\,cm Pearson-Readhead (PR) survey (Pearson \& Readhead
1981), which had already been started on the Global VLBI Network in
the late 1970s, had been extended into the first and second
Caltech-Jodrell Bank surveys (CJ1: Polatidis et al.\ 1995, CJ2: Taylor
et al.\ 1994). In order to allow unified use of these samples, we
defined the Caltech-Jodrell Flat-spectrum (CJF) sample of 293 sources
(Taylor et al.\ 1996). The present paper is a brief status update on
the effort to measure the motions in the jets of all of the CJF
sources.

\section{CJF}

CJF is a complete flux-limited sample of 293 flat-spectrum radio
sources, drawn from the 6\,cm and 20\,cm Green Bank Surveys. The CJF
sample is now 97\% optically identified, and spectroscopic redshifts
are available for 92\% of the sources. Most of them are quasars, at
redshifts ranging from $z=0.263$ to $z=3.886$. There are also a
significant number of low redshift objects, but in addition there are
some 25 galaxies and 20~BL\,Lac objects at $z>$0.6; enough to allow a
meaningful comparison of their motion statistics with that of quasars
of the same redshift and luminosity. Between March 1990 and November
2000 there were numerous multi-antenna global VLBI and VLBA snapshot
runs at 6\,cm. For every source, including those in the PR subsample,
this has resulted in 3--5 ``modern'' VLBI datasets, spanning 4--8
years.

\begin{figure}[htb!]
\begin{center}
\vspace{-10pt}
\includegraphics[clip,height=0.9\textwidth,angle=-90]{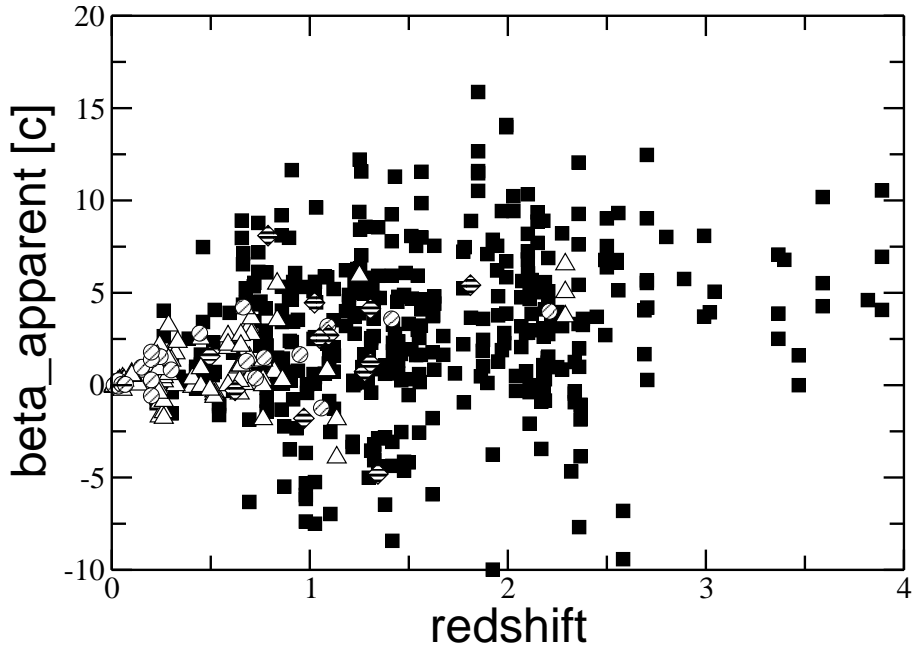}
\vspace{-15pt}
\end{center}
\caption{Overview of all of the motions as currently measured, for 597
   components in 262 CJF sources, calculated with
   $H_\circ=65$ km\,s$^{-1}$\,Mpc$^{-1}$ and $q_\circ=0.5$. Squares are
   quasars, circles are BL\,Lac Objects, triangles are galaxies, and
   diamonds are unclassified optical hosts.}
\end{figure}

\section{Measuring jet component motions}

All observing epochs for all sources are being analysed in the same
standardised way (e.g., Britzen et al.\ 1999, 2001). We show in Figure~1
all of the currently available motions for all of the source
components, using $H_\circ=65$\,km\,s$^{-1}$\,Mpc$^{-1}$ and
$q_\circ=0.5$. The source structures can usually be described
fairly well by using a limited number of Gaussian image components,
fitted to the $(u,v)$-data using {\sc difmap}. However, it is not trivial to
derive meaningful, consistent motions from the positions of these
Gaussians.

The flux density and indeed the morphology of a feature clearly
recognisable at one epoch may have changed considerably in the next
epoch. This can lead to difficulties in ``connecting'' features between
epochs. Expansion or brightness profile changes in resolved knots can
also affect the centroid positions obtained. Furthermore, while in
numerous cases the ``core'' is an unmistakable and presumably fixed
reference point for the positions of the other features, there are also
sources in which new jet features were emerging from the core (often
accompanied by radio flaring), and these events can lead to a temporary
shift in the ``core'' centroid. In some sources there is no component
which, by its compactness and/or spectral index, is clearly the
non-moving core; evidently, sometimes there is either little radio
emission produced in the immediate vicinity of the central engine, or,
more likely, the core emission is faint at 5\,GHz due to either
synchrotron self-absorption or free-free absorption.

We think that most and quite possibly all of the negative motions
(i.e., inward towards the core) in Figure~1 are due to component
misidentifications or to brightness changes which do not necessarily
correspond to motion of a well-defined component. It would obviously be
of great interest to find and then study a case of significant inward
motion, but so far none of the candidates has survived close scrutiny.

Some of the components also clearly accelerate, decelerate, or follow a
curved trajectory; these are not well characterised by a single
separation velocity from the core. We plan to make a proper statistical
description of the non-linear jet motions encountered, which we hope
will prove to be very valuable in confronting the various models for
jet acceleration, collimation, and propagation, and (shock) models for
the generation of radio emitting particles inside those jets.

We have found that many of the ambiguities are substantially
ameliorated for those sources where 4 or 5 rather than 3 epochs are
available. However, there are no indications of any systematic velocity
differences relating to the number of epochs, the time intervals
between the epochs, or the total time spanned between the first and the
last epoch. The measurements shown in Figure~1 also do not show selection
effects, due to either resolution limitations (against low velocities at
high redshifts), or to time sampling limitations (against high
velocities at low redshifts).

\section{Results}

We are still in the process of defining an adequately homogeneous
sub-sample of the highest quality motion measurements in which the
impact of the effects described above is minimised. But a number of
interesting conclusions can already be drawn from the current motion
dataset. This includes 262 sources (the other 41 CJF sources have
either no redshift or no jet components that could be monitored), with
a total of 597 components.

The quasars have a mean apparent velocity of $2.9c$. This is also close
to the peak in the distribution, with lower apparent velocities being
somewhat less common, as are higher values; most of the quasar apparent
velocities are below $8c$, but the tail of the distribution extends up
to $16c$. These statistics confirm the finding of Vermeulen \& Cohen
(1994) that many of the superluminal velocities reported in the first
studies of the 1970s and 1980s have turned out to be amongst the
fastest ones in the larger population now available. Apart from the
selection effect that faster motions may have been more amenable to
study early on, many of the very brightest sources are also amongst the
most luminous ones, and we also confirm the strong correlation between
apparent velocity and 5\,GHz luminosity (see below) already noted in
Vermeulen (1995).

There is a pronounced difference between the apparent velocity
distributions of the quasars on the one hand, and the galaxies and BL\,Lac 
objects on the other hand. Whereas the mean apparent velocity in the
quasar jets is $2.9c$, the galaxies have a mean apparent velocity of
$0.9c$, and in fact here the distribution peaks at the lowest values,
$\le1c$, with most of the apparent velocities below $2c$, although
there is a tail in the distribution which goes all the way up to
$10c$. The BL\,Lacs may be slightly faster than the galaxies on average,
with a mean apparent velocity of $1.4c$, but again the distribution
peaks at the lowest values, $\le1c$. Most of the BL\,Lacs have apparent
velocities below $4c$, and the distribution then cuts off steeply,
with no values larger than $6c$. In general, such results are in line
with previous work (e.g., Vermeulen \& Cohen 1994), and with the
predictions of beaming and unification models. But we have a much more
extensive sample than hitherto available, with different source types
at comparable redshifts and luminosities. Before coming to final
conclusions, we are carefully re-examining the optical
identifications. For example, using very deep optical images some
objects may have been labelled as galaxies on the basis of faint
extended emission which would not have been visible in less sensitive
images. We also have the impression that it is important for the final
statistics to differentiate not only between broad-lined (quasar-like)
and narrow-lined (galaxy-like) objects, but also between between
strong-lined and weak-lined (FR-I BL Lac ?) objects.

Doppler favouritism in simple beaming models combines with the solid
angle available to predict a sample in which the apparent velocity
distribution is sharply peaked at the maximum possible value
($\beta\gamma$); such models would predict that the measurements shown
in Figure~1 would crowd against an upper velocity envelope (e.g.,
Vermeulen \& Cohen (1994).  This is obviously not the case, and the
actual apparent velocity distribution may instead be dominated by a
distribution of Lorentz $\gamma$ factors which spans a very broad range
over the sample of radio sources (see also Vermeulen \&\ Cohen
1994). There is evidence for this in the very strong correlation with
luminosity of the mean apparent velocity, which was already found in a
subset of the CJF dataset (Vermeulen 1995), in the sense that no high
velocities exist at low luminosity. The correlation can be attributed
to a form of Malmquist bias in a sample selected on beamed radio
emission (e.g., Lister \& Marscher 1997).

Perhaps it will prove to be possible to turn this relationship around,
and to deduce, from the observed distributions of apparent velocity,
luminosity, and redshift, the luminosity distribution of the parent
population from which the CJF sample arose through beaming. But, we
first have to study carefully whether the Lorentz factors responsible
for the selection of the sources, through relativistic boosting of the
radio cores, are indeed tightly related to the Lorentz factors
corresponding to the motions of the components. The latter could reflect
patterns or shocks moving through the fluid of the jets. And even if
the components move at the bulk flow velocity, bending of the jets will
give different apparent speeds.

By compiling the statistics of the apparent velocities as a function of
distance from the core, we also plan to investigate whether there is
evidence for jet acceleration or deceleration between the core and the
moving components. 
Perhaps related to this, very interesting hints have emerged at this
Green Bank workshop that the average apparent velocity observed might
decrease when going to lower observing frequency.
The
apparent velocities for CJF at 5\,GHz, reported above, seem to be about
a factor of two slower than those reported in these proceedings for the
15\,GHz VLBA Survey (Zensus et al., 
these proceedings, page 27),
and those in turn might be
another factor of two lower than the 43\,GHz velocities reported at this
workshop by Marscher (these proceedings, page 133).
Most of these measurements are as
yet preliminary, and the differences between the selections of the
various samples will also have to be carefully understood, before it
will be warranted to incorporate correlations between observing
frequency and apparent velocity firmly into the models. But several
options do spring to mind immediately. Jet bending, or deceleration
along the jets might be involved. Alternatively, each jet might
incorporate a range of flow velocities, possibly in a ``spine-sheath''
or ``onion-skin'' geometry (perhaps akin to some models for FR-I
sources, for instance in Laing et al.\ 1999). A decrease in average
apparent velocity with decreasing frequency might then result from an
increase of the average size and/or distance from the core of the
components observed when going to lower frequency. In turn this could
plausibly be connected to changing turnover frequencies for features
with synchrotron self-absorbed spectra.

And so we have again found at this workshop that Ken Kellermann plays a
central role in superluminal motion studies, in which a novel feature
now is that we will be able to compare two large surveys: the 6\,cm CJF
and the 2\,cm VLBA Survey, which Ken initiated, and in which he provides
important momentum and inspiration.

\printindex

\begin{references}

\reference Britzen, S., et al.
1999, 
in ASP Conf. Ser. Vol. 159, BL Lac Phenomenon, ed.  L. O. Takalo, 
A. Sillanp\"a\"a, (San Francisco: ASP), 431 

\reference Britzen, S., et al.
2001, 
in IAU Symp. 205, Galaxies and Their Constituents at the Highest Angular Resolutions, ed.  R. T. Schilizzi, S. Vogel, F. Paresce, \& M. Elvis (San Francisco: ASP), 106 

\reference Cohen, M. H., et al.
1971, 
ApJ, 170, 207 

\reference Cohen, M. H., et al. 1977, 
Nature, 268, 405 

\reference Kellermann, K. I. 1966, 
ApJ, 146, 621 

\reference Kellermann, K. I. et al. 1968, 
ApJ, 153, L209 

\reference Kellermann, K. I., et al.
1973, 
ApJ, 183, L51 

\reference Kellermann, K. I., et al.
1974, 
ApJ, 189, L19 

\reference Kellermann, K. I. 1985, 
   Com. Ap., 11, 69 

\reference Kellermann, K. I., Vermeulen, R. C., Zensus, J. A., \&
   Cohen, M. H. 1998, 
AJ, 115, 1295 

\reference Laing, R. A., Parma, P., de Ruiter, H. R., \& Fanti, R. 1999, 
MNRAS, 306, 513--530.

\reference Lister, M. L., \& Marscher, A. P. 1997, 
ApJ, 476, 572 


\reference Pauliny-Toth, I. I. K., \& Kellermann, K. I. 1966, 
ApJ, 146, 634 

\reference Polatidis, A. G., et al.
1995,
ApJS, 98, 1 

\reference Rees, M. J. 1967, 
   MNRAS, 135, 345 

\reference Shaffer, D. B., Cohen, M. H., Jauncey, D. L., \& Kellermann,
   K. I. 1972, 
ApJ, 173, L147 

\reference Taylor, G. B., et al.
1994, 
ApJS, 95, 345 

\reference Vermeulen, R. C. 1995, 
Proc. Natl. Acad. Sci. USA, 92, 11385 
(correction in Proc. Natl. Acad. Sci. USA, 93, 6846) 

\reference Vermeulen, R. C., \& Cohen, M. H. 1994, 
ApJ, 430, 467 


\end{references}
\end{document}